\documentclass[sigconf]{acmart}
\usepackage{multirow}
\usepackage{graphicx}
\usepackage{subcaption}
\usepackage{amsmath}
\usepackage{tikz}
\usepackage{balance}

\AtBeginDocument{%
  \providecommand\BibTeX{{%
    \normalfont B\kern-0.5em{\scshape i\kern-0.25em b}\kern-0.8em\TeX}}}

\copyrightyear{2023} 
\acmYear{2023} 
\setcopyright{acmlicensed}
\acmConference[KDD '23]{Proceedings of the 29th ACM SIGKDD Conference on Knowledge Discovery and Data Mining}{August 6--10, 2023}{Long Beach, CA, USA}
\acmBooktitle{Proceedings of the 29th ACM SIGKDD Conference on Knowledge Discovery and Data Mining (KDD '23), August 6--10, 2023, Long Beach, CA, USA}
\acmPrice{15.00}
\acmDOI{10.1145/3580305.3599758}
\acmISBN{979-8-4007-0103-0/23/08}

\begin{document}

\title{A Collaborative Transfer Learning Framework for Cross-domain Recommendation}

\author{Wei Zhang}
\affiliation{%
\institution{Meituan}
\city{Beijing}
\country{China}
}
\email{zhangwei180@meituan.com}

\author{Pengye Zhang}
\affiliation{%
\institution{Meituan}
\city{Beijing}
\country{China}
}
\email{zhangpengye@meituan.com}

\author{Bo Zhang}
\affiliation{%
\institution{Meituan}
\city{Beijing}
\country{China}
}
\email{zhangbo126@meituan.com}

\author{Xingxing Wang}
\affiliation{%
\institution{Meituan}
\city{Beijing}
\country{China}
}
\email{wangxingxing04@meituan.com}

\author{Dong Wang}
\affiliation{%
\institution{Meituan}
\city{Beijing}
\country{China}
}
\email{wangdong07@meituan.com}

\renewcommand{\shortauthors}{Wei Zhang, Pengye Zhang, Bo Zhang, Xingxing Wang, \& Dong Wang}

\begin{abstract}

In the recommendation systems, there are multiple business domains to meet the diverse interests and needs of users, and the click-through rate(CTR) of each domain can be quite different, which leads to the demand for CTR prediction modeling for different business domains. The industry solution is to use domain-specific models or transfer learning techniques for each domain. The disadvantage of the former is that the data from other domains is not utilized by a single domain model, while the latter leverage all the data from different domains, but the fine-tuned model of transfer learning may trap the model in a local optimum of the source domain, making it difficult to fit the target domain. Meanwhile, significant differences in data quantity and feature schemas between different domains, known as domain shift, may lead to negative transfer in the process of transferring. To overcome these challenges, we propose the Collaborative Cross-Domain Transfer Learning Framework (CCTL). CCTL evaluates the information gain of the source domain on the target domain using a symmetric companion network and adjusts the information transfer weight of each source domain sample using the information flow network. This approach enables full utilization of other domain data while avoiding negative migration. Additionally, a representation enhancement network is used as an auxiliary task to preserve domain-specific features. Comprehensive experiments on both public and real-world industrial datasets, CCTL achieved SOTA score on offline metrics. At the same time, the CCTL algorithm has been deployed in Meituan, bringing 4.37\% CTR and 5.43\% GMV lift, which is significant to the business.
\end{abstract}





\maketitle

\section{Introduction}
In online recommendation systems, traditional CTR prediction models\cite{covington2016deep}\cite{sun2022graph}\cite{guo2017deepfm}\cite{zhou2019deep}\cite{wang2017deep}\cite{zhou2018deep} focus on a specific domain, where the CTR model serves only a single business domain after being trained with samples collected from this domain. At large e-commercial companies like Alibaba and Amazon, there are often many business domains that need CTR prediction to enhance user satisfaction and improve business revenue. Since different business domains have overlapping user groups and items, there exist commonalities among these domains \cite{zhu2021cross}. Enabling information sharing is beneficial for learning the CTR model of each domain. Therefore, how to use cross-domain transfer learning to leverage information from the rich domains to help the poor domains has become one of the main research directions in the industry.

The common cross-domain modeling scheme in the industry mainly falls into two paradigms\cite{sheng2021one}\cite{yang2022adasparse}\cite{ganin2016domain}\cite{zhu2019dtcdr}\cite{ouyang2020minet}\cite{li2021dual}\cite{richardson2007predicting}: 1) union and mix up the source and the target samples, and then perform multi-task learning techniques to boost the performance in all domains; 2) pre-train a model using the mixed or the data-abundant source domain data, and then finetune it in the data-deficient target domain to fit the new data distribution. In the first approach, the domain-specific and domain-invariant characteristics are learned through different types of network designs, where the domain indicator is commonly used to identify domains. In the fine-tuning paradigm, the approach mainly argues that the data in the target domain is not enough to make the parameters to be fully trained. Therefore, in such opinions, the pre-train is necessary to train the parameters first, and then make the model converge to the optimum through the target domain data. The two paradigms have been proven to be effective in some scenarios, however, they may still fall short on some occasions, as we will discuss these limitations in the following.

For the multi-task learning solutions, all the source domain data is mixed up with the target ones, and they make the assumption that the model architecture could definitely identify the difference and similarities. However, this may be too idealistic. The user behaviors and item groups may be different as the domain changes, and the amount of data among domains can vary. Therefore, the training process can be easily dominated by data-abundant domains, resulting in insufficient training in sparse domains (i.e., \textbf{seesaw effect}\cite{zhang2021survey}\cite{crawshaw2020multi}\cite{xie2022multi}). As a result, such approaches are not friendly to sparser target domains. 

For the pre-train and fine-tune solutions, the fine-tuning process is expected to make use of the trained parameters and guide the optimization through target samples. However, the optimal solution trained in the source domain may be a \textbf{local minimum} for the target domain(i.e., the non-optimal solution finetune problem\cite{kumar2022fine}\cite{he2021analyzing}). The data distribution shift widely exists among domains, e.g., the click-through rate of the same item is different when it is displayed in different creative forms on different domains. When the parameters are well-trained to fit the source distribution, it is hard for the model to jump out and find a new suitable optimum in the target domain. Therefore, it is necessary to evaluate how much beneficial information the source can bring to the target.

In order to solve such problems caused by source domain samples in cross-domain modeling, we propose the CCTL (\textbf{C}ollaborative \textbf{C}ross-domain \textbf{T}ransfer \textbf{L}earning framework) algorithm. The CCTL algorithm mainly includes three parts: the Symmetric Companion Network, Information Flow Network, and Representation Enhancement Network. The symmetric companion network trains the mixed model (the target and source samples) and the pure model(only target samples). According to the difference in the effect of the two parts, it is evaluated whether the current source domain samples are helpful to the target domain.
The information flow network transfers the sample weight calculated for each source sample and performs semantic alignment between domains. Finally, the representation enhancement network act as an auxiliary task to maintain the domain-specific characteristics in each domain.

The main contributions of this paper are as follows:

\begin{itemize}
\item We propose CCTL, a simple but effective cross-domain modeling framework. CCTL can select samples from the source domain that are beneficial to the target domain training and add them to the target domain training, reducing the introduction of invalid samples and noise samples.
\item In order to evaluate the gain efficiency of information flow from the source domain to the target domain, we propose an Information Flow Network to evaluate the potential gain of each source domain sample to the target domain.
\item We propose the Representation Enhancement Network, through contrastive learning, to allow the id embedding of the source domain and the target domain to accommodate as much different information as possible.
\item We evaluate CCTL on the industrial production dataset and deploy it in the display advertising system of Meituan in 2022. The consistent superiority validates the efficacy of CCTL. Up to now, the deployment of CCTL brings 4.37\% CTR and 5.43\% GMV lift.
\end{itemize}

\section{METHODOLOGY}

In this section, we first introduce a general expression formulation for cross-domain CTR prediction. Next, we provide an overview of CCTL, explaining our approach broadly. Afterward, we delve into the details of each component through several subsections.

\subsection{Problem Setup}
We define the cross-domain CTR prediction problem as using data from one or multiple source domains to enhance the performance of the CTR model on the target domain. Specifically, we annotate an input sample $X_i^d$ and the corresponding label information $y_i^d \in \{0, 1\}$ from domain $d \in \{s, t\}$, where $s$ represents the source domain and $t$ represents the target domain. The goal of cross-domain CTR prediction is to train a model using CTR samples from two business domains, namely the source domain sample set $(X_i^s, y_i^s)$ and the target domain sample set $(X_j^t, y_j^t)$, to predict the click probability in the target domain. 

The objective of the cross-domain CTR prediction problem can be formulated as equation \eqref{formulation:J}, shown as follows:

\begin{gather}
    \begin{aligned}
        \mathcal{J} = minimize \left \{  \frac{1}{N_t}\sum_{x_i \in X^t}^{N_t} Loss(y_i^t, f(x_i, \Theta_{model}|X_j^s, X_j^t)) \right \}
    \end{aligned}
    \label{formulation:J}
\end{gather}
where $x_i$ is a sample in the target sample set $X^t$, $N_t$ is the number of target samples, $y_i^t$ is the label information of the target domain, $f$ is the model's function form, with input containing samples from both source and target domains ($X^s$ and $X^t$) and model parameters $\Theta_{model}$, $Loss$ is the loss function over the target sample set $X^t$ with $N_t$ samples, where cross-entropy is commonly adopted in the CTR estimation problem.

\subsection{Model Overview (CCTL)}
\begin{figure*}[htb]
  \centering
  \includegraphics[width=0.7\linewidth]{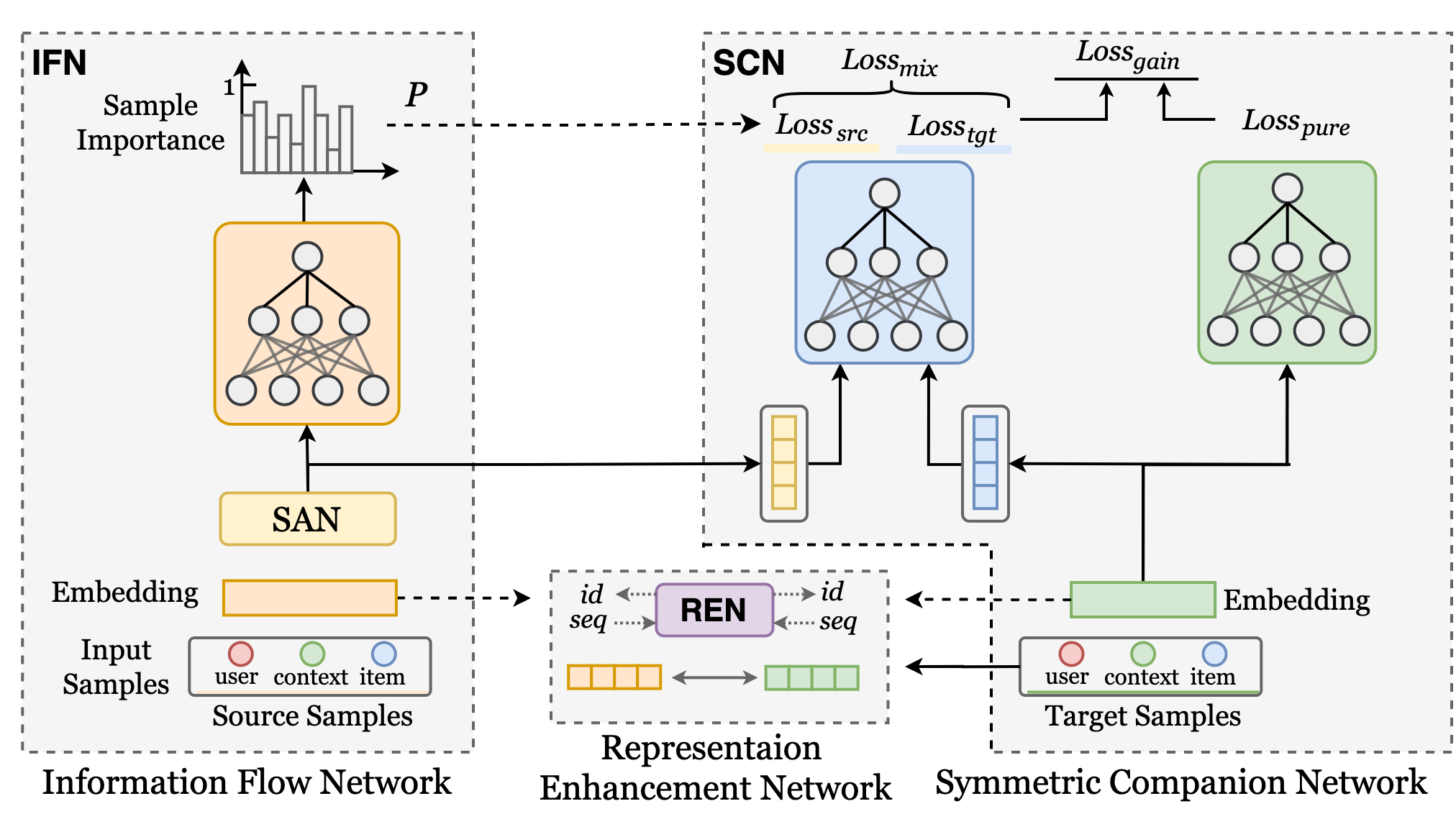}
  \Description{fig.2}
  \captionsetup{skip=5pt}
  \caption{An overview of the CCTL model. a)Symmetric Companion Network(SCN) evaluates the information gain from the source domain samples to the target domain through the dual-tower framework. b)Information Flow Network(IFN) evaluates how much information can be brought by each source sample, and help the information to flow partially through an output weight. c) Representation Enhancement Network(REN) is an auxiliary component that tries to preserve the unique information of different domains through a contrastive design.}
  \label{fig:CCTL}
  \vspace{-1em}  
\end{figure*}

The overall workflow of CCTL is shown in Figure \ref{fig:CCTL}. It is composed of three components: the Symmetric Companion Network (SCN), the Information Flow Network (IFN), and the Representation Enhancement Network (REN). The main purpose of SCN is to evaluate the information gain from the source domain samples to the target domain through the dual-tower framework. The role of IFN is to evaluate how much information can be brought by each source sample, and help the information to flow partially through an output weight. And the semantic field of source and target domains is also aligned by IFN. REN is an auxiliary component that tries to preserve the unique information of different domains through a contrastive design. The three components work together to make the best use of cross-domain information and optimize the model training.

\subsection{Symmetric Companion Network}

\textbf{Main Description}: Train the target domain network and detect negative transfer from the source domain. By designing a symmetrical structure, more useful information from source domain samples can be transferred to the target domain.


The source and target domains have similarities in terms of users and items, but their data distributions differ. There may be information noise from the source domain that can negatively impact the target domain, causing a phenomenon known as the negative transfer. Simple transfer learning techniques like pre-training and fine-tuning or multi-task learning by combining all samples from both domains may not result in significant improvement, as the impact of source domain samples on the target domain is not evaluated. 
The main purpose of SCN is to identify negative transfer during training, and its principle will be explained in detail. The evaluation of source domain samples is primarily carried out by IFN, which will be covered in Section \ref{sec:ifn}. 

\begin{figure}[h]
  \centering
  \includegraphics[width=0.9\linewidth]{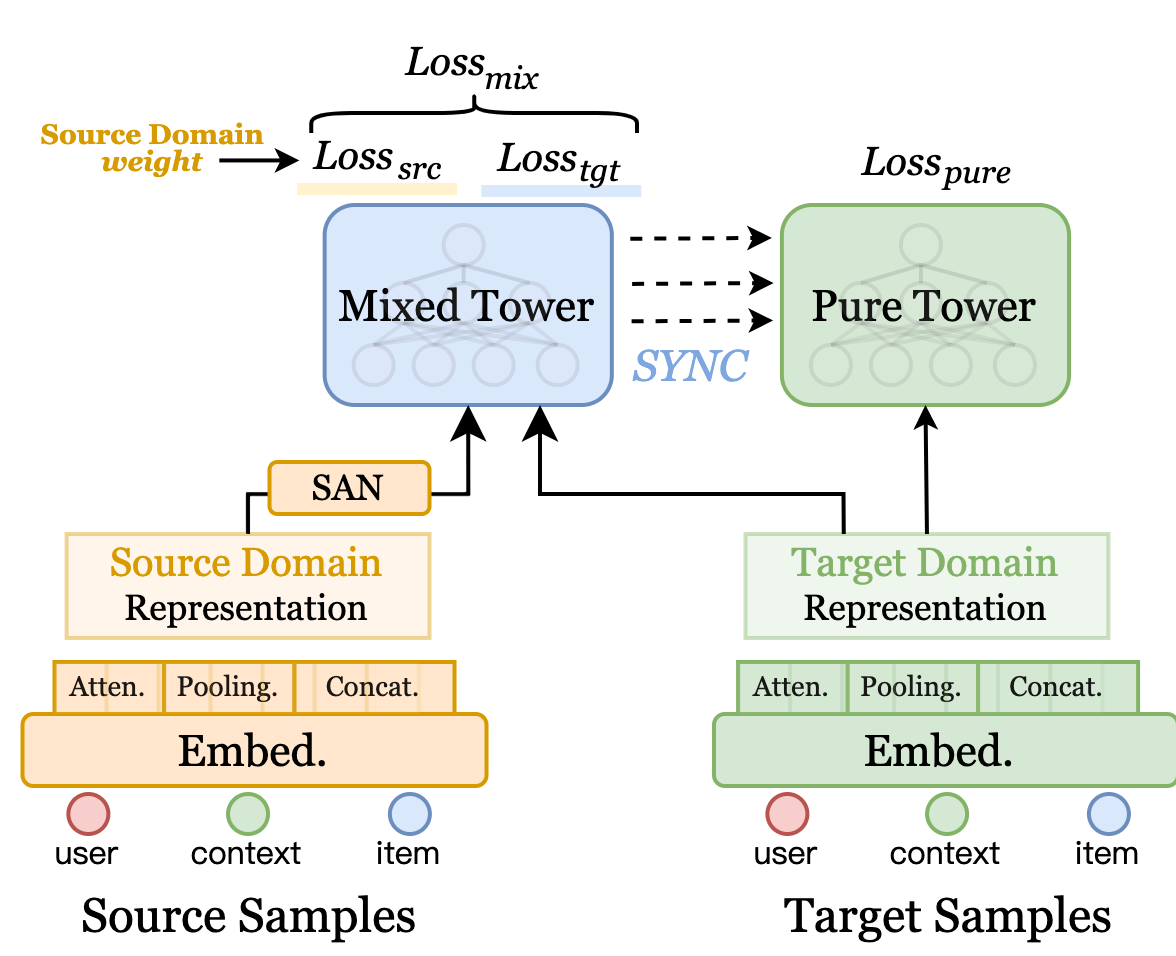}
  \Description{An illustration of the SCN component.}
  \caption{An illustration of the SCN component.}
  \label{fig:scn}
  \vspace{-2em}  
\end{figure}

To accurately evaluate the impact of the source samples on the target domain, the straightforward approach is to compare the performance of a model trained on mixed domain samples to the one trained solely on the target domain. Then, the influence can be calculated as the difference in offline metrics between the two models. The SCN designed based on this concept adopts a dual-tower network architecture as shown in Figure \ref{fig:scn}. In SCN, one tower, referred to as the mixed tower, receives inputs from both the source and target domains simultaneously, while the other tower, referred to as the pure tower, only receives inputs from the target domain. In the view of the principle of the control variables, the difference between the mixed and pure towers is solely attributed to the impact of the source domain.

\vspace{-1em}
\subsubsection{Feature Embedding}
we encode features in different fields into one-hot encodings and map them to dense, low-dimensional embedding vectors suitable for deep neural networks. The attention mechanism is applied later for the embedded item id and user behavior sequence(item ids). We omit the attention mechanism details because it's not the key design in this paper, and can be replaced by other modeling strategies such as a sequential encoder, etc. Finally, the sample of the target domain is transformed into a vector representation $\textbf{v}_i^t$, shown as follows:
\begin{gather}
    \begin{aligned}
        \textbf{v}_i^t = [\textbf{e}_1 || \textbf{e}_2 || ... || \textbf{e}_{n_t} ]
    \end{aligned}
\end{gather}
where $n_t$ is the number of categorical features in the target domain, and $||$ denotes the vector concatenation operation. Similarly, we apply the same embedding techniques for the samples in the source domain.

It is worth noting that in real scenarios, there will be differences in the feature scheme of the source and target domains, which will cause the shapes of $\textbf{v}^t$ and $\textbf{v}_i^s$ to be inconsistent, thus preventing them from using the same network for training. This paper will use the semantic alignment network(SAN) structure in the information flow network(IFN) to process the Embedding of $\textbf{v}^s$ into the same shape as $\textbf{v}^t$. See the IFN chapter in section \ref{sec:ifn} for more details. 

\subsubsection{Mixed Tower}
As shown in Figure \ref{fig:scn}, the mixed tower will have two parallel data inputs, i.e., reading source domain samples and target domain samples simultaneously for training. In this way, the loss of the mixed tower is calculated as follows:
\begin{gather}
\begin{aligned}
Loss_{mixed} &= Loss_{src} + Loss_{tgt} \\
&= { \sum_{x_i \in X^s}^{N_s} \textbf{p}_i^s L(y_i^s, f(x_i, \Theta _{mixed}))} \\
&+ { \sum_{x_i \in X^t}^{N_t} L(y_i^t, f(x_i, \Theta _{mixed}))}
\end{aligned}
\end{gather}
where $x_i$ is the sample from the source or the target domain, $f$ is short for the neural network operation, $L$ denotes the loss function, and we use cross-entropy in this paper, and $\Theta_{mixed}$ is the trainable parameters in the mixed network. $\textbf{p}_i^s$ is the output of another component indicating the weight of each source sample, and we will discuss it in the next subsection. In this way, we can obtain a mixed tower trained jointly by the source domain and the target domain.

\subsubsection{Pure Tower}
The pure tower only reads the target samples, which are exactly the same as the target samples in the mixed tower. and the loss in the pure target domain is formulated as follows:
\begin{gather}
Loss_{pure} = { \sum_{x_i \in X^t}^{N_t} L(y_i^t, f(x_i, \Theta_{pure}))}
\end{gather}
where $x_i \in X^t$ is the sample only from the target domain, and $\Theta_{pure}$ is the parameters in the pure tower. $L_{pure}$ guides the back-propagation of the pure network during training, which is the same as the traditional single-domain training process. In this way, we can obtain a pure tower trained only by the target domain.

\subsubsection{Impact of Source Domain}
 In SCN, the two towers (mixed and pure) have the same shape of network structure, learning rate, etc. The only difference is that the mixed tower additionally reads the samples of the source domain, and these source domain samples affect the network parameters in the mixed tower through the gradient of backpropagation. Therefore, we only need to use the same set of target domain samples to calculate the loss on the two towers, and the difference between the two losses is the impact result of the source domain. Fortunately, during the training process, the loss on the same target domain samples has already been calculated as $Loss_{pure}$ and $Loss_{tgt}$, so the information gain can be defined as follows:
\begin{gather}
\begin{aligned}
r &= Loss_{gain} = Loss_{pure} - Loss_{tgt} \\
&= {\sum_{x_i \in X^t}^{N_t} [L(y_i^t, f(x_i, \Theta _{pure})) - L(y_i^t, f(x_i, \Theta _{mixed}))]}
\end{aligned}
\end{gather}
where $N_t$ is the number of target samples, $\Theta _{pure}$ is the training parameter of the pure tower in SCN. The term $r$ represents how much the loss reduces after the parameters are updated by additional source samples. When the information brought by the source domain samples is positive, the mixed tower in the SCN can predict more accurately. In particular, when predicting the same target domain samples, $Loss_{tgt}$ will be smaller than $Loss_{pure}$, and $r>0$ at this time. Otherwise, the noisy information will result in $r<=0$, meaning that there is a negative transfer in the cross-domain training.

\subsubsection{Parameter Synchronization}
To reduce the offset caused by the two-tower learning route, we perform parameter synchronization every $k$ steps, where $k = 1000$ in this paper, that is, synchronize the parameters of the mixed network to the pure network, so that $Loss_{mixed/src/tgt/pure}$ will not introduce too much noise due to network learning offset.

\subsection{Information Flow Network}
\label{sec:ifn}

\textbf{Main Description}: Receive the reward and gradient updates from the SCN, and transmit source domain sample weights and representations. Since not all information from source domain samples is useful, so weighted transmission is necessary.

The SCN is capable of assessing negative transfer and detecting it during cross-domain training. As mentioned before, negative transfer in existing methods is often caused by misusing source domain samples, some of which are only partially useful for the target domain. For instance, a user's interests in movies and books may overlap, but not completely match due to differences in presentation format.

\begin{figure}[h]
  \vspace{-1em}  
  \centering
  \includegraphics[width=0.85\linewidth]{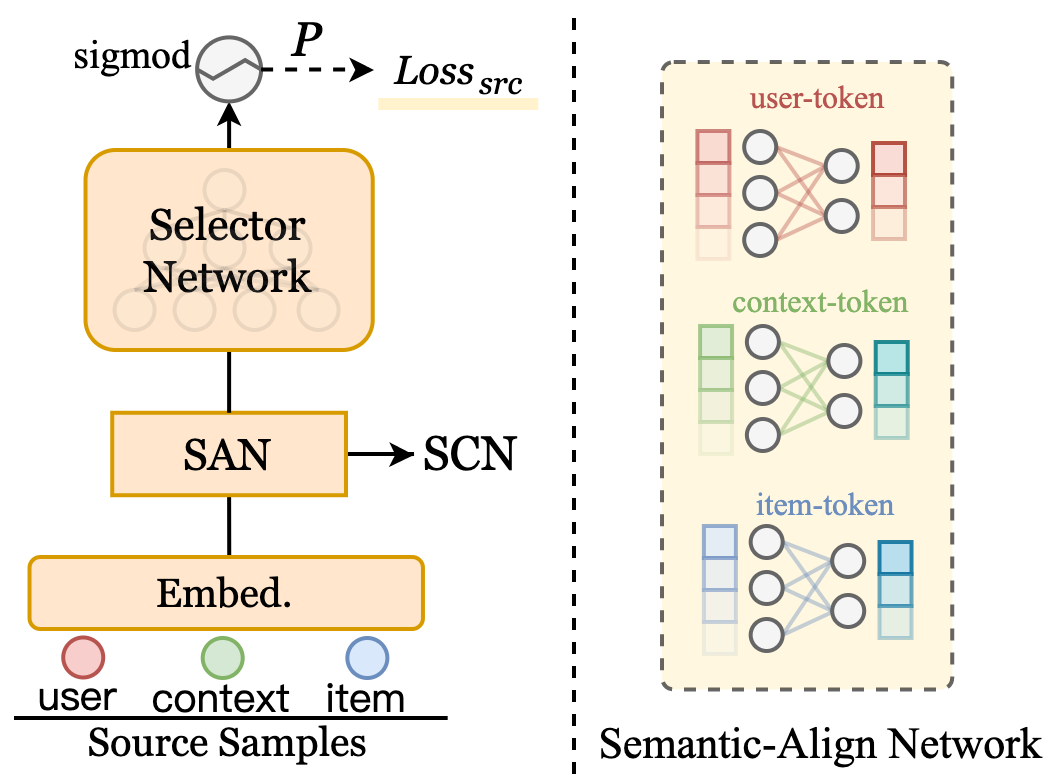}
  \Description{An illustration of the IFN component.}
  \caption{An illustration of the IFN component.}
  \label{fig:ifn}
  \vspace{-1em}  
\end{figure}

The main function of the information flow network(IFN) in figure \ref{fig:ifn} has three aspects: 1) Evaluate the potential profit coefficient of a single sample in the source domain for the target domain.2) Consistent alignment of the evaluation goal with the goal of maximizing the effect of the target domain model.
3) Effective transfer of the information gained in the source domain to the target domain.

Note that IFN mainly predicts the benefit of a single source domain sample to the target domain, while SCN accurately evaluates the total benefit of the source domain based on the difference in the prediction accuracy of the two towers for the target domain. Conceptually, IFN is responsible for prediction and SCN is responsible for evaluation. In section \ref{sec:select network}, the role of SCN in helping IFN training will be explained in detail. In the following section \ref{sec:Semantic-Align Network}, we will first discuss how to effectively transfer the information obtained in the source domain to the target domain.

\subsubsection{Semantic-Align Network}
\label{sec:Semantic-Align Network}
In this section, we will discuss the challenge of addressing the discrepancy in sample representation caused by the dissimilarities of features in different domains. In cross-domain transfer learning, it is crucial to effectively transfer information from the source domain to the target domain, however, the mismatch of features can create a significant difference in sample representation. The number of features in various domains can be vastly different, leading to the discrepancy between the source and target domains in the SCN. Hence, a single mixed network cannot be utilized for samples from both domains.

Despite the challenges, transfer learning has proven to be a powerful tool in the fields of computer vision (CV) and natural language processing (NLP). The success of transfer learning is attributed to its ability to preserve cross-domain information through semantic tokens between different domains. In the field of CV, a token can be represented as pixels composing a point, line, or surface, while in NLP, a token can be represented as a word. When facing a new scenario, even though the problem form may change, these basic tokens remain highly consistent with the source task, allowing for knowledge transfer and application.

We consider users, items, and contexts as basic semantic tokens, meaning that users interact with items in specific contexts. This user-item-context relationship can be applied to any recommendation scenario. To handle this, we split the source domain sample vector representation $\textbf{v}_i^s$ into three separate tokens: $\textbf{v}_{user}^s$, $\textbf{v}_{item}^s$, and $\textbf{v}_{context}^s$. Each token comprises features related to it. For example, the user token representation can be simplified as follows:
\begin{gather}
    \begin{aligned}
    \textbf{v}_{user}^s = [\textbf{e}_{uid} || \textbf{e}_{male} || \textbf{e}_{ios}]
    \end{aligned}
\end{gather}
where $\textbf{e}_{uid}$, $\textbf{e}_{male}$, $\textbf{e}_{ios}$ are the embeddings of user-id, gender, and device type respectively.

In order to achieve the alignment of the source domain to the target domain, we design a special compression network. The input shape of this network is equivalent to the source domain semantic token, and the output shape is equivalent to the corresponding target domain semantic token:
\begin{gather}
    \begin{aligned}
\textbf{v}_{user}^{s '} = MLP(\textbf{v}_{user}^{s}) \in \mathbb{R}^{dim^t_{user}}
    \end{aligned}
\end{gather}
where $dim^t_{user}$ is the dimension of embedded user features in the target domain, and $MLP$ is a set of DNN layers with activation functions. The transformed source sample representations have the same dimensionality as the target domain and are aligned at the semantic token granularity. In this way, the subsequent source domain can use the same network as the target domain for later processing.

But it has to be mentioned that if in an ideal multi-domain environment, the source domain and the target domain use exactly the same features, then this part of the SAN structure can be ignored, that is, the source domain can be directly used without additional semantic token mapping. The original embedding result of the domain is enough.

\subsubsection{Selector Network}
\label{sec:select network}

 The selector network is mainly responsible for evaluating the information gain of the source domain samples for the target domain. 
Regarding the network design, this paper uses a multi-layer MLP network, with the last layer using the sigmoid activation function:
\begin{gather}
\begin{aligned}
    \textbf{p}_i^s = \frac {1} {1 + exp{(- W_m h_{m-1} + b_m)}}
\end{aligned}
\end{gather}
where $m$ is the number of the DNN layers, $h_{m-1}$ is the output of the last DNN layer, and $W_m$ and $b_m$ are the params to be trained.

In the previous SAN structure, the source domain sample $\textbf{v}_i^s$ has been obtained and fed into the dual-tower to obtain the final information gain evaluation.

Through the weight $\textbf{p}_i^s$ of IFN, the loss on each source sample is dynamically adjusted. The network parameters in SCN are thus updated through gradient from both the target domain and weighted source domain, so the information from the source domain can be "\textbf{partially adapted}" to the target. 

The selector network itself has no explicit labels, i.e., there is no label information to indicate whether a source sample is suitable for the target or not. Recall that in the SCN structure, it is already possible to evaluate the gain increase $r$ of a batch of source domain samples for the target domain, but this gain $r$ is a scalar value, which cannot correspond to each source domain sample one-to-one, and thus cannot use the traditional stochastic gradient descent to get updated. But fortunately, there are already mature solutions to such problems in the field of reinforcement learning, which can be updated through reinforcement learning algorithms.

The term $r_k$ can be regarded as the reward, where $k$ is the batch indicator. The accumulated reward can be defined as $r^{accu}=r_k + \gamma * r_{k+1}  + ...  + \gamma ^{n-k} * r_{k}$, where $r_n$ is the reward of the last batch and $\gamma$ is the weight factor and is set to 0.80 uniformly.

Finally, we adopt REINFORCE\cite{williams1992simple} algorithm to update the IFN component, and the parameters are updated as follows:
\begin{gather}
    \begin{aligned}
    \Theta_{ifn} \leftarrow \Theta_{ifn}  + \alpha \frac{1}{N} \sum_{i=1}^{N_s} \triangledown \Theta_{ifn} log(\textbf{p}_i^s) r^{accu}
    \end{aligned}
    \label{formulation:REINFORCE}
\end{gather}
where $\Theta_{ifn}$ is the parameter of the IFN, $\textbf{p}_i^s$ is the output weight for the $i$th source sample, $r^{accu}$ is the accumulated reward, and $\alpha$ controls how much influence the source samples should have on the target domain. Like this, the $r^{accu}$ is regarded as the indirect loss part. In order to make the gradient update of the selector more stable, we accumulate the reward $r^{accu}$ and perform the gradient return of the IFN every 1000 steps, so as to avoid excessive fluctuation of the batch effect and stabilize the parameter update process.

With the symmetric design of SCN, we are able to evaluate the information gain of source domain samples. When the information gain brought by the sample is positive, the mixed tower in SCN can predict the target more accurately, so that $Loss_{tgt}$ decreases, and the reward is positive, that is, the reward encourages IFN to increase this weight on such source samples. Similarly, when the sample has a negative impact on the target domain, $Loss_{tgt} > Loss_{pure}$, the reward is negative, and the selection strategy of IFN will be corrected when the gradient is updated.

\subsection{Representation Enhancement Network}

\textbf{Main Description}: Use contrastive learning strategies to distinguish the similarity and differences between samples from different domains, so as to reduce the harmful impact of domain-specific knowledge in the source domain on the target domain.

In the paradigm of transfer learning, we have found through practice that a single embedding representation cannot fully cover cross-domain meaning when the differences between domains are large. Because each domain has its own unique knowledge, even for the same item in different domains.
For example, in the search and feed scenarios, the search domain has query information, while the feed domain does not. 
Therefore, we attempt to retain some domain-specific knowledge during the transfer process to accommodate situations where the source and target domains have significant differences. Specifically, in our scenario, our approach is mainly based on the following two ideas:

 \textbf{Maximizing the similarity between sequence embedding.} For the same user, although there are differences in the behavior sequences in different domains, the user's interest remains stable and is expressed through different heterogeneous sequence representations in different domains. Therefore, we believe that the user representations between different domains should be kept as consistent as possible. User representations mainly consist of sequence features, so we believe that it is necessary to maximize the similarity between sequence embeddings.
 
 \textbf{Minimizing the similarity between ID embedding.} Based on maximizing user similarity, we also want to retain the unique information of each domain, expecting to provide differentiated information from the source domain to the target domain. ID Embedding is a microscopic representation of a domain, so we try to control the differentiated representation of the same ID between different domains, thereby retaining more additional information.
 
\begin{figure}[h]
  \vspace{-1em}
  \centering
  \includegraphics[width=0.9\linewidth]{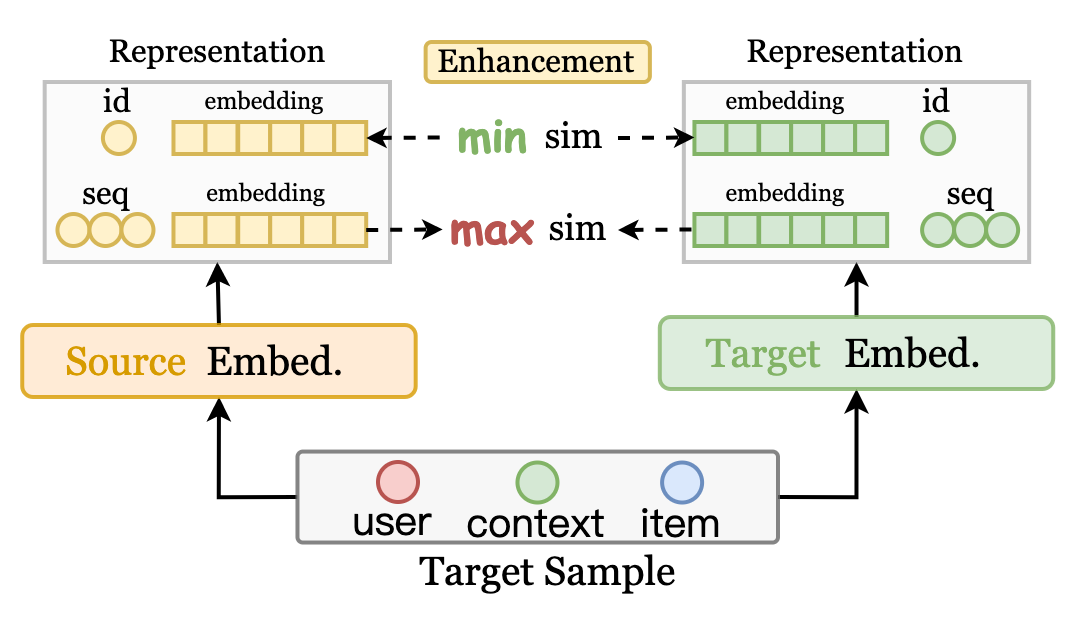}
  \Description{An illustration of the REN component.}
  \captionsetup{skip=5pt}
  \caption{An illustration of the REN component.}
  \label{fig:ren}
  \vspace{-1em}  
\end{figure}

As illustrated in Figure \ref{fig:ren}, we proposed the Representation Enhancement Network (REN) to address the above ideas.
These two ideas act as constraints, somewhat similar to minimax game design, can be formalized as follows:

\begin{equation}
    \begin{aligned}
        & min \left[ sim(\textbf{v}_{id}^{s} , \textbf{v}_{id}^{t}) \right ] \\
        & max \left[ sim(\textbf{v}_{seq}^{s}, \textbf{v}_{seq}^{t} ) \right ]
    \end{aligned}
\end{equation}
where $\textbf{v}_{id}^{s}$ is the id embedding after source domain embedding lookup, and $\textbf{v}_{id}^{t}$ is the target one. $\textbf{v}_{seq}^{s}$ is for the sequence embedding. In particular, for the sequence embedding, we use the average pooling for the user behavior id sequence. $sim$ is the similarity metric, where we use the normalized cosine similarity for simplicity as is shown:
\begin{gather}
    \begin{aligned}
        sim(\textbf{v}_i, \textbf{v}_j) = \frac {\textbf{v}_i \cdot \textbf{v}_j}{||\textbf{v}_i||_2 \cdot ||\textbf{v}_j||_2}
    \end{aligned}
\end{gather}
where ${||\cdot||_2}$ is the length in Euclidean space for the vector. The term evaluates how similar are the two vectors. The loss formula can be expressed as follows through an equivalent transformation while optimizing both max and min at the same time:

\begin{gather}
    \begin{aligned}
        Loss_{ren} = sim(\textbf{v}_{id}^{s}, \textbf{v}_{id}^{t}) - sim( \textbf{v}_{seq}^{s}, \textbf{v}_{seq}^{t} )
    \end{aligned}
\end{gather}

The auxiliary loss of REN is added to the SCN and acts as an assistant to help the mixed tower better recognize the domain-specific and invariant features. The main idea of the dual embedding design is to maintain the differences between domains, and also, the separation of embedding also avoids possible conflicts of different encoding among domains.

\section{EXPERIMENTS}

In this section, extensive offline and online experiments are performed on both the large-scale recommender system in Meituan and public benchmark datasets to answer the following research questions:

\textbf{RQ1} Does our proposed method outperform the baseline methods?

\textbf{RQ2} How does each part of our CCTL model work?

\textbf{RQ3} How does the model perform when deployed online?

Before presenting the evaluation results, we first introduce the experimental setup, including datasets, baselines, metrics, and parameter settings.

\subsection{Experimental Setup}
\subsubsection{\textbf{Datasets.}}
We adopt public datasets and industrial datasets to comprehensively compare CCTL models and baseline models. The statistics of the datasets are shown in Table \ref{tab:dataset}.

\textbf{Amazon dataset}: This dataset has been widely used to evaluate the performance of collaborative filtering approaches. We use the two largest categories, Books and Movies\&TV, to form the cross-domain datasets. We convert the ratings of 4-5 as positive samples. The dataset we used contains 979,151 user-book ratings and 432,286 user-movie ratings. There are 61,437 shared users, 835,005 books, and 368,953 movies. We aim to improve the movie-watching domain recommendation(as the target domain led from knowledge). The statistics are summarized in the table, and hence we hope to improve the target domain performance by transferring knowledge from source domains.

\textbf{Taobao dataset}: This dataset was first released by Alibaba-Taobao and is widely used as a common benchmark in CTR prediction tasks. It is a user behavior log for Taobao mobile application, including click, purchase, add-to-cart, and favorite behaviors. This dataset involves 987,994 users and 4,162,024 items. We rank according to the number of samples under each category, divide the top 70\% categories as the source domain, and the remaining categories as the target domain.

\textbf{Industrial dataset}: This dataset is an industrial dataset collected by Meituan App which is one of the top-tier mobile Apps in our country. It is much larger than the Amazon and Taobao public datasets. In the source domain, the data contains the sample information of the user on the list page, involving 230M users and 1.1M items, where $M$ is short for $10^6$. The target domain mainly includes 183M users and 0.7M items. There is an intersection between the two domains on users and items, which are 67.83\% and 70.26\% respectively.

\begin{table*}[ht]
  \caption{\textcolor{black}{Statistics of public and industrial datasets.}}
  \label{tab:dataset}
  \centering
  \setlength{\abovecaptionskip}{0.cm}
  \setlength{\belowcaptionskip}{-0.cm}
  \begin{tabular}{c|ccc|ccc}
    \toprule
    \multirow{2}{*}{Datasets} & \multicolumn{3}{c}{Source Domain} & \multicolumn{3}{c}{Target Domain} \\
    ~ & Users & Items & Samples & Users & Items & Samples \\
    \midrule
    Amazon & 61,437 & 835,005 & 979,151 & 61,437 & 368,953 & 432,286 \\
    Taobao & 690,006 & 69,386,796 & 70,000,525  & 296,716 & 29,436,098 & 30,150,807 \\
    Meituan & 230M & 1.1M & 1200M & 183M & 0.7M & 250M \\
  \bottomrule
\end{tabular}
  \vspace{-1em}  
\end{table*}

\subsubsection{\textbf{Baselines}}
We compare both single-domain and cross-domain methods. Existing cross-domain methods are mainly proposed for cross-domain recommendation and we extend them for cross-domain CTR prediction when necessary (e.g., to include attribute features rather than only IDs and to change the loss function).

A. Single-Domain Methods.
\begin{itemize}
\item LR. Logistic Regression\cite{richardson2007predicting}. It is a generalized linear model.
\item DNN. Deep Neural Network\cite{cheng2016wide}. 
It contains an embedding layer, a fully connected layer(FC), and an output layer.
\item DeepFM. DeepFM model\cite{guo2017deepfm}. It combines factorization machine(FM, the wide part) and DNN (the deep part), thus modeling multi-order feature interactions.
\item DIN. Deep Interest Network model\cite{zhou2018deep}. It models dynamic user interest based on historical behavior sequences.
\item AFM. Attentional Factorization Machine\cite{xiao2017attentional}. It learns the weight of feature interactions through an attention design.
\end{itemize}

B. Cross-Domain Methods.
\begin{itemize}
\item Finetune. Finetune Network\cite{yang2022click}. A typical transfer learning method in the field of CTR prediction trains on the source domain and finetunes the networks in the target domain.
\item DANN. Domain-Adversarial Neural Network\cite{ganin2016domain}. It maps the source and the target domain to the same space through the adversarial network and a gradient flip design, so that the domain-invariant information can be learned.
\item CoNet. Collaborative cross Network\cite{hu2018conet}. The cross-connection units are added on MLP++ to enable the knowledge to transfer among domains.
\item MiNet. Mixed Interest Network\cite{ouyang2020minet}. It jointly models long and short-term interest in the source domain with a short-term interest in the target domain.
\item STAR. An industrial framework using star topology design\cite{sheng2021one}, which models different domains of CTR as multi-task learning.
\item CCTL. CCTL is the proposed approach in this paper.
\end{itemize}

\subsubsection{\textbf{Parameter Settings}}
We set the dimension of the embedding vectors for each feature as embedding dim = 8. We set the number of fully connected layers in neural network-based models as L=4, where the industrial dataset is [1024, 512, 128, 1] and the public dataset is [256, 128, 32, 1]. For the industrial dataset, the batch size is set to 8192, while set to 128 for the Amazon and Taobao datasets. All the methods are implemented in Tensorflow and we use Adam\cite{duchi2011adaptive} as the optimization algorithm uniformly. Each method has been run 3 times and the reported are the average results.

\subsubsection{\textbf{Metrics}}
We evaluate the CTR prediction performance with two widely used metrics. The first one is the area under the ROC curve (AUC) which reflects the pairwise ranking performance between click and non-click samples. The other metric is log loss, which is to measure the overall likelihood of the test data and has been widely used for classification tasks. At the same time, We deploy models online in real industrial systems. Our online evaluation metric is the real CTR($CTR=\frac{\#click}{ \#pages}$) which is defined as the number of clicks over the number of impressions and GMV($GMV=1000 * \frac{\#pay \, amount}{\#pages}$) which represents the revenue amount per thousand impressions, as an evaluation metric in our A/B tests.

\vspace{-1em}
\subsection{RQ1: Does our CCTL model outperform the baseline model?}
We evaluate the performance of CCTL and baseline models on three datasets. The performances of single-domain and multi-domain are compared simultaneously. The single-domain model is introduced for two purposes: 1) to explain the limitations of the current single-domain, and show the non-optimal fintune problem; 2) to make an intuitive comparison with the multi-domain model. From Table \ref{experimental result table}, we have the following observations:

\begin{table*}[ht]
  \caption{Experimental results on different methods.}
  \label{experimental result table}
  \begin{tabular}{cccccccc}
    \toprule
    \multirow{2}{*}{Domains} ~ & ~& \multicolumn{2}{c}{Meituan} & \multicolumn{2}{c}{Amazon} & \multicolumn{2}{c}{Taobao} \\
    \cline{3-8} 
    ~ & Model & AUC & Loss & AUC & Loss & AUC & Loss\\
    \midrule
    \multirow{5}{*}{Single Domain} & LR & 0.6805 & 0.5589& 0.7421 & 0.4677 & 0.7907 & 0.3894\\
    ~ & DNN & 0.6876 & 0.5537& 0.7456 & 0.4448 & 0.7953 & 0.3741\\
    ~ & DeepFM & 0.6893 & 0.5497& 0.7473 & 0.4432 & 0.7962 & 0.3724 \\
    ~ & DIN & 0.6932 & 0.5468 & 0.7480 & 0.4454 & 0.8095 & 0.3692\\
    ~ & AFM & 0.6923 & 0.5457& 0.7460 & 0.4492 & 0.7990 & 0.3854\\
    \hline
    \multirow{6}{*}{Cross Domain} & Finetune & 0.6921 & 0.5452& 0.7564 & 0.4381 & 0.8280 & 0.3585 \\
    ~ & DANN & 0.6946 & 0.5404 & 0.7610 & 0.4331 & 0.8420 & 0.3601 \\
    ~ & CoNet & 0.6942 & 0.5416 & 0.7621 & 0.4332 & 0.8418 & 0.3666 \\
    ~ & MiNet & 0.6953 & 0.5377 & 0.7665 & 0.4305 & 0.8447 & 0.3640 \\
    ~ & STAR & 0.6945 & 0.5391 & 0.7672 & 0.4298 & 0.8440 & 0.3643 \\
    ~ & CCTL & \textbf{0.7001} & \textbf{0.5288}& \textbf{0.7790} & \textbf{0.4230} & \textbf{0.8521} & \textbf{0.3578}\\
  \bottomrule
  \end{tabular}
  \vspace{-1em}  
\end{table*}

In single-domain, the more complex the model, the better the effect, but its marginal returns show diminishing status. 
At the same time, we noticed that as the computational complexity of the model increases, the benefit of AUC increases to a marginally diminishing state, especially when both DIN and AFM adopt variants based on the attention structure, and the performance improvement is limited. Our analysis may be based on the current model complexity, which is gradually approaching the amount of information contained in the sample, resulting in limited effect improvement.

For cross-domain methods, the overall effect is better than single-domain ones, and the benefits of cross-domain information flow are better than the increase in the complexity of a single model. However, the finetune model may not be better than the traditional single model (for example, on the Meituan dataset, its effect basically slightly worse than DIN and AFM), we think the reason is the non-optimal solution finetune problem mentioned above. CoNet introduces cross-connection units to enable dual knowledge transfer across domains. DANN maps the source and target domains to the same space through an adversarial network and gradient flip design so that the samples in the source domain similar to the target can be integrated. However, these two methods also introduce higher complexity and random noise, because all samples are used indiscriminately. The MiNet method starts from the perspective of information flow from the source domain to the target domain, tries to split information from a long-term and short-term perspective, and transfers interest information from the source domain through a specific network structure. But the domain-specific info is not considered, and it lacks the ability to filter and constrain the information on the source domain. 
The STAR method adopts separate MLPs to tackle multiple domains, while we see the negative transfer in Meituan and Taobao datasets, which is probably caused by domain shifts.
In contrast, our proposed CCTL use an information flow network to provide filtered and beneficial samples to the target domain. Apart from the alignment for semantic tokens, we further use a representation enhancement network to maintain the domain differences. Therefore, the performance in the target domain can be improved compared with baselines.

\subsection{RQ2: How does each part of our CCTL model work?}

\subsubsection*{\textbf{A.Effect of the Selector Network}}

\begin{figure}[h]
  \setlength{\abovecaptionskip}{10pt}
  \begin{subfigure}{.5\linewidth}
    \centering
    \includegraphics[width=\linewidth]{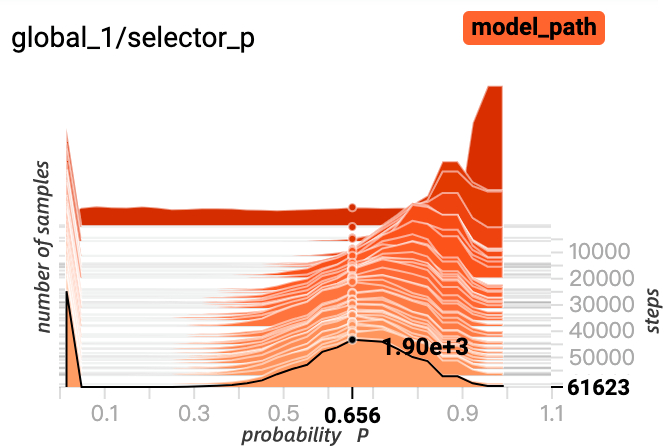}
    \caption{Selector $P$'s trend in training.}
    \label{fig:sub1}
  \end{subfigure}%
  \begin{subfigure}{.5\linewidth}
    \centering
    \includegraphics[width=\linewidth]{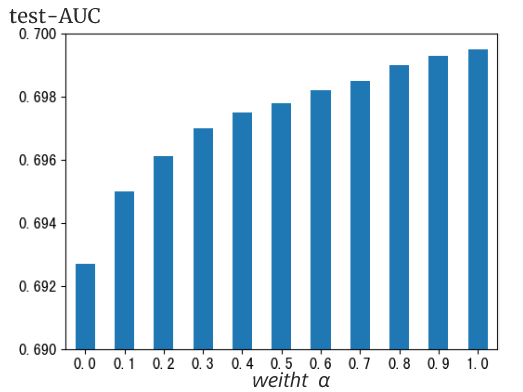}
    \caption{Sensitivity for selector weight.}
    \label{fig:sub2}
  \end{subfigure}
  \caption{The effect evaluation of the Selector Network.}
  \label{fig:esn}
  \vspace{-1em}  %
\end{figure}

We evaluate the role of the Selector Network in IFN in two ways on the Meituan dataset. Firstly, we track the output sample weight P of the selector during the training process, then plot the distribution histogram in tensorboard. The horizontal axis is the selector output probability, the vertical axis is the number of samples, and the right side is marked as the steps of model training. As can be seen from Figure \ref{fig:esn}(a), the distribution changes when training, and finally most of the samples form a similar Gaussian distribution with an average of P=0.656. And some of the source domain samples are not selected (P=0 on the left). This shows that the probability of the selector network is gradually learned as the model is trained.

\vspace{-0.5em}  
Secondly, we adjust the weight parameter $\alpha $ in eq \eqref{formulation:REINFORCE} of the selector network, where $\alpha \in [0, 1]$. By adjusting $\alpha $ the importance of the selector network can be changed. When $\alpha $ is 0, it degenerates into an ordinary multi-domain sample task. As can be seen from Figure \ref{fig:esn}(b), when the size of $\alpha$ is changed from small to large, the effect test-AUC in the target domain shows a growing trend, and as $\alpha$ continues to approach 1, the effect growth slows down. This shows the selector has a positive influence on the model.

\subsubsection*{\textbf{B.Effect of different modules.}}
In the proposed CCTL framework of this paper, IFN and REN can be viewed as auxiliary tasks to assist the main module SCN in training the target domain. Therefore, in order to verify their impact, we conducted ablation experiments on the IFN and REN modules respectively. The results of the ablation experiments are shown in table \ref{tab:ren}.

\begin{table}[h]
  \setlength{\abovecaptionskip}{0.cm}
  \setlength{\belowcaptionskip}{-0.cm}
  \caption{\textcolor{black}{The ablation study for the components in CCTL.}}
  \label{tab:ren}
  \centering
  \begin{tabular}{ccc}
    \toprule
     Ablations & AUC & LogLoss\\
    \midrule 
     CCTL w/o IFN & 0.8385 & 0.3647 \\
     CCTL w/o REN & 0.8493 & 0.3602 \\
     Full CCTL & 0.8521 & 0.3578  \\
  \bottomrule
\end{tabular}
  \vspace{-1em}  
\end{table}

It can be seen from table \ref{tab:ren} that the effect of removing IFN is greater than that of releasing REN. Removing IFN means using all source domain samples for training. In this case, the model's performance is only slightly better than that of the Finetune, which indicates that using source domain samples indiscriminately is not a wise choice. When the REN module is removed, the representations of different domains may be close to each other, resulting in a lack of discrimination between different domains. Through the above ablation experiments, the necessity of IFN and REN has been demonstrated, and it is shown that IFN contributes more benefits, indicating that the selection and weighting of source domain samples are very important.

\subsection{RQ3: How does the model perform when deployed online?}

\begin{figure}[h]
  \centering
  \includegraphics[width=0.9\linewidth]{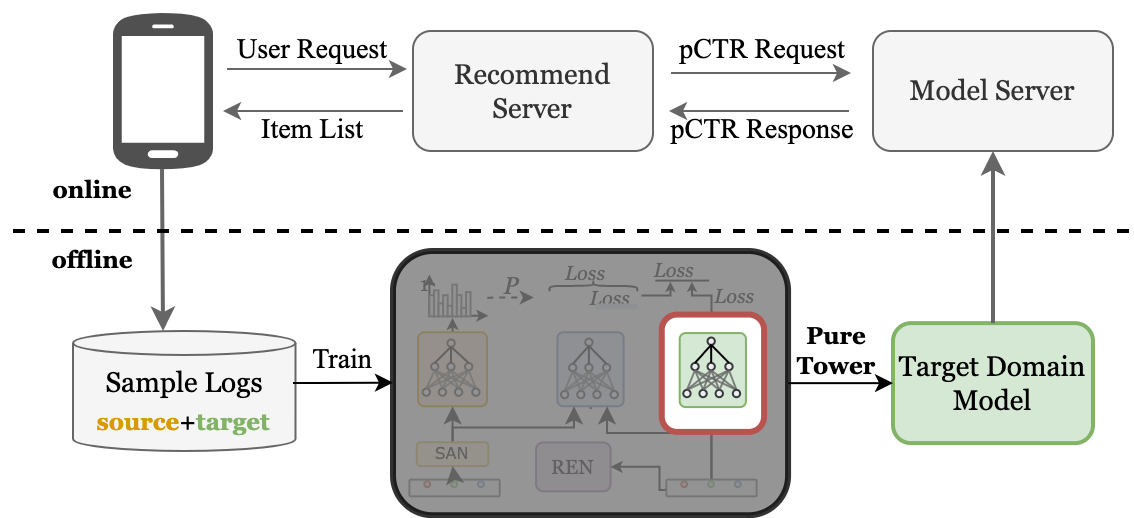}
  \caption{Architecture of the online deployment with CCTL.}
  \label{online deployment picture}
  \vspace{-2em}  
\end{figure}

We deployed CCTL in Meituan App, where the ad serving system architecture is shown in Figure \ref{online deployment picture}. It is worth noting that although we involved multiple network structures during offline training when serving online, we only need to use the pure tower in SCN(note: the parameters in the mixed tower have been synchronized to the pure tower during training), i.e., just export the network of target domain in the online model service. In this way, we ensure that its model complexity is equivalent to that of the online model, without adding additional calculations to the online service.

We conducted online experiments in an A/B test framework over three months during Sep.2022 - Dec. 2022, where the base serving model is a variant of DIN according to our business characteristics. Our online evaluation metric is the real CTR($CTR=\frac{\#click}{\#pages}$) which is defined as the number of clicks over the number of impressions and GMV($GMV=1000 * \frac{\#pay \, amount}{\#pages}$) which represents the revenue amount per thousand impressions. A larger online CTR indicates the enhanced effectiveness of a CTR prediction model. The online A/B test shows that CCTL leads to an increase of online CTR of 4.37\% and GMV of 5.43\% compared with the base model. This result demonstrates the effectiveness of CCTL in practical CTR prediction tasks. After the A/B test, CCTL serves the main ad traffic in Meituan App.

\section{RELATED WORKS}
Our paper belongs to the field of cross-domain CTR prediction and has close ties to single-domain CTR prediction. In this section, we provide a brief overview.

\textbf{Cross-Domain CTR Prediction:} Cross-domain learning refers to a technique that enhances the performance of a target domain by incorporating knowledge from source domains. This is accomplished through various algorithms, including transfer learning\cite{niu2020decade}, multi-task learning\cite{zhang2021survey}, and multi-view learning\cite{yan2021deep}. These methods aim to transfer knowledge in a deep manner, allowing multiple base networks to benefit from each other, and provide more flexible representation transfer options. Cross-domain learning is useful for various applications, including computer vision, natural language processing, and recommendation systems, where data and tasks across domains are related. In the CTR prediction field, STAR\cite{sheng2021one} tries to mix multiple sources of data for training a unified model. CoNet\cite{hu2018conet} aims to make cross-domain information flow more efficiently by enabling dual knowledge transfer through cross-mapping connections between hidden layers in two base networks. MiNet\cite{ouyang2020minet} further divides multiple sources into long-term and short-term interests and directly models cross-domain interests. Under the paradigm of union training, the DASL\cite{li2021dual} model includes both the dual embedding and dual attention components to jointly capture cross-domain information. Meanwhile, AFT\cite{hao2021adversarial} learns feature translations between different domains within a generative adversarial network framework.

Data sparsity\cite{man2017cross} is also a classic problem that Cross-Domain Recommendation seeks to solve. This may be due to the sparsity of the target domain data itself, or to local sparsity caused by the long-tail effect and cold-start effect. Commonly used methods for such problems involve using data from the source domain to address the sparsity of target domain data. For example, ProtoCF\cite{sankar2021protocf} efficiently transfers knowledge from arbitrary base recommenders by extracting meta-knowledge to construct discriminative prototypes for items with very few interactions. PTUPCDR\cite{zhu2022personalized} proposes that a meta-network fed with users' characteristic embeddings is learned to generate personalized bridge functions for achieving personalized transfer of user preferences. TLCIOM\cite{krishnan2020transfer} uses domain-invariant components shared between dense and sparse domains to guide neural collaborative filtering to achieve information flow from one dense domain to multiple sparse domains. CCDR\cite{xie2022contrastive} and SSCDR\cite{kang2019semi} address the problem of the cold-start domain lacking sufficient user behaviors through techniques such as contrastive learning and semi-supervised learning.

All these approaches tried different ways of separating domain info, or integrating the source samples through different model architectures, while few of them focus on the information weight of the source sample, which is the key point this paper addresses.

\textbf{Single-Domain CTR Prediction:} CTR prediction models have shifted from traditional shallow approaches to modern deep approaches. These deep models commonly use embedding and MLP. Wide\&Deep\cite{cheng2016wide} and DeepFM\cite{guo2017deepfm} improve expression power by combining low- and high-order features. 
User behavior is transformed into low-dimensional vectors through embedding techniques. DIN\cite{zhou2018deep} uses attention to capture the diversity of user interest in historical behavior. DIEN\cite{zhou2019deep} adds an auxiliary loss and combines attention with GRU to model evolving user interest. MIND\cite{li2019multi} and DMIN\cite{xiao2020deep} argue for multiple representations to capture complex patterns in user and item data, using capsule networks and dynamic routing. The Transformer is used for feature aggregation inspired by the success of self-attention in sequence-to-sequence learning. MIMN\cite{pi2019practice} employs a memory-based architecture to aggregate features and addresses long-term user interest modeling. SIM\cite{pi2020search} extracts user interest using cascaded search units, improving scalability and accuracy in modeling lifelong behavior data. The framework of this paper focuses on the information flow design, so these single-domain techniques can be merged with ours easily in the SCN design, which may further boost the model performance.

\vspace{-\baselineskip}
\section{CONCLUSION}
In this paper, we propose a collaborative cross-domain transfer learning framework, which enables the reuse of beneficial source domain samples. In the framework, the information flow network(IFN) can evaluate the importance of the source domain samples, in which the SAN helps align meaningful semantic tokens among domains. The symmetric companion network(SCN) is designed with a symmetric structure to approximately fit the information gain of the samples in the source domain to the target domain. The representation enhancement network(REN) maintains domain-specific characteristics through a contrastive learning mechanism. In this way, only beneficial information in the source domain is transferred to the target, which reduces the seesaw effect and the negative transfer. Then we validate our method on both public and real-world industrial datasets by performance comparison and ablation study, which proved its effectiveness. Finally, we deployed the CCTL model on Meituan App. The CCTL model has brought significant business improvement and served as the mainstream traffic.


\bibliographystyle{ACM-Reference-Format}
\balance
\bibliography{adfp607}
\end{document}